\documentclass[11pt,oneside,english,reqno,a4paper]{amsart}

\usepackage{mathptmx}
\usepackage{graphicx}
\usepackage{latexsym}
\usepackage{float}
\usepackage{diagbox}
\usepackage{caption}
\usepackage{multicol}
\usepackage[section]{placeins}
\usepackage{mathrsfs}
\usepackage[numbers]{natbib}
\usepackage[fit]{truncate}
\usepackage{amssymb,amsmath,amsfonts}
\usepackage[english]{babel}
\usepackage[utf8]{inputenc}
\usepackage[compatible]{algpseudocode}
\usepackage{hhline}

\newcommand{\by}{\boldsymbol{y}}
\newcommand{\bI}{\boldsymbol{I}}
\newcommand{\bu}{\boldsymbol{u}}
\newcommand{\bbet}{\boldsymbol{\beta}}
\newcommand{\bgam}{\boldsymbol{\gamma}}
\newcommand{\beps}{\boldsymbol{\varepsilon}}
\newcommand{\bx}{\boldsymbol{x}}
\newcommand{\bz}{\boldsymbol{z}}
\newcommand{\bX}{\boldsymbol{X}}
\newcommand{\bZ}{\boldsymbol{Z}}
\newcommand{\bF}{\boldsymbol{F}}
\newcommand{\bC}{\boldsymbol{C}}
\newcommand{\bP}{\boldsymbol{P}}
\newcommand{\bH}{\boldsymbol{H}}
\newcommand{\bS}{\boldsymbol{S}}
\newcommand{\bmin}{\textbf{-}}
\newcommand{\bone}{\boldsymbol{1}}
\newcommand{\bQ}{\boldsymbol{Q}}
\newcommand{\btet}{\boldsymbol{\vartheta}}
\newcommand{\btau}{\boldsymbol{\tau}}
\newcommand{\bphi}{\boldsymbol{\varphi}}
\newcommand{\bet}{\boldsymbol{\eta}}
\newcommand{\pp}{\vphantom{\beta}}

\newcommand{\grb}{\texttt{grbLMM} }
\newcommand{\grba}{\texttt{grbLMM}$^a$ }
\newcommand{\grbb}{\texttt{grbLMM}$^b$ }

\newenvironment{titlemize}[1]{%
	\paragraph{#1}
	\begin{itemize}}
	{\end{itemize}}

\DeclareMathOperator{\Var}{\text{Var}}

\DeclareMathOperator{\argmin}{arg\,min}

\DeclareMathOperator{\dg}{\text{diag}}

\numberwithin{equation}{section}

\makeatletter
\def\author@andify{%
	\nxandlist {\unskip ,\penalty-1 \space\ignorespaces}%
	{\unskip {} \@@and~}%
	{\unskip \penalty-2 \space \@@and~}%
}
\makeatother

\begin{document}

\title{Gradient Boosting for Linear Mixed Models}
\author{Colin Griesbach}
\author{Benjamin S\"afken}
\author{Elisabeth Waldmann}
\thanks{The work on this article was supported by the DFG (Projekt WA 4249/2-1) and the Volkswagen Foundation. Colin Griesbach performed the present work in partial fulfilment of the requirements for obtaining the degree ‘Dr. rer. biol. hum.’ at the Friedrich-Alexander-Universität Erlangen-N\"urnberg.}

\address{Department of Medical Informatics, Biometry, and Epidemiology, Friedrich-Alexander-Universität Erlangen-Nürnberg, Waldstr. 6, D-91054 Erlangen}
\email{colin.griesbach@fau.de}
\address{Chair of Statistics, Georg August University, Humboldtallee 3, D-37073 G\"ottingen}
\email{benjamin.saefken@uni-goettingen.de}
\address{Department of Medical Informatics, Biometry, and Epidemiology, Friedrich-Alexander-University Erlangen-Nürnberg, Waldstr. 6, D-91054 Erlangen}
\email{elisabeth.waldmann@fau.de}

\begin{abstract} Gradient boosting from the field of statistical learning is widely known as a powerful framework for estimation and selection of predictor effects in various regression models by adapting concepts from classification theory. Current boosting approaches also offer methods accounting for random effects and thus enable prediction of mixed models for longitudinal and clustered data. However, these approaches include several flaws resulting in unbalanced effect selection with falsely induced shrinkage and a low convergence rate on the one hand and biased estimates of the random effects on the other hand. We therefore propose a new boosting algorithm which explicitly accounts for the random structure by excluding it from the selection procedure, properly correcting the random effects estimates and in addition providing likelihood-based estimation of the random effects variance structure. The new algorithm offers an organic and unbiased fitting approach, which is shown via simulations and data examples.
\end{abstract}

\maketitle

\section*{Introduction}
Linear mixed models \citep{Laird.1982} are a popular modelling class for longitudinal or clustered data. Due to their comparatively simple structure and high fle\-xibility, these models are widely used, for instance in clinical surveys with repeated measurements. However, the application of these models can go far beyond that and include any kind of hierarchical modelling and models with penalized smooth components, see \citep{Anderssen.1974} and \citep{Wahba.1985} or \citep{Wood.2017} for an overview.

Inference in mixed effects models is mostly based on maximum likelihood or restricted maximum likelihood and there are well-known software packages available for the estimation of these kind of models \citep{Bates.lme4, Pinheiro.2020}. Furthermore, classical methods for inference such as tests \citep{Crainiceanu.2004} and model selection criteria \citep{Vaida.2005, Greven.2010} have been developed for mixed models. While all of the approaches above offer a convenient framework for inference with a modest number of parameters, models with a high number of parameters are covered by different regularization approaches. In \citep{Schelldorfer.2011, Groll.2014, Hui.2017} various lasso \citep{Tibshirani.1996, Friedman.2010} penalties have been adapted to mixed models and in \citep{Bradic.2020} p-Values are constructed for high-dimensional mixed model settings using Neyman orthogonalization. Another popular approach to regularized regression are various boosting methods.

Boosting originally evolved from the field of machine learning as an approach to classification problems \citep{Freund.1996} and was later adapted to statistical models \citep{Breiman.1998, Breiman.1999, Friedman.2000}. The com\-ponent-wise and iterative fitting scheme of boosting algorithms offers advantages like implicit variable selection, improved prediction quality and makes the method applicable for high dimensional data. While achieving similar results to classical approaches (e.g. lasso) at simple setups like linear models, boosting outperforms those methods when proceeding to complex models, since the modular system allows for generic predictors with covariates of all forms \citep{Hepp.2016}. A broad overview of boosting methods can be found in \citep{Mayr.2014} as well as in \citep{Buehlmann.2007} with implementations in the \texttt{mboost} package \citep{Hothorn.mboost} available on \texttt{CRAN}.

A framework for random effects was added to the \texttt{mboost} syntax \citep{Kneib.2009, Hofner.2014} bringing the advantages of boosting to analysis of longitudinal and clustered data. However, several specifications of this modelling framework lead to irregular selection processes and biased estimates like the following example shows.

For $n = 1,\dots,50$ clusters and $j = 1,\dots,10$ observations per cluster consider a basic linear mixed model simulation setup
\begin{equation}\label{eq_example}
	y_{ij} = \beta \cdot x_i + \gamma_{i} + \varepsilon_{ij}, \quad \gamma_{i} \sim \mathcal{N}(0, \tau^2)
\end{equation}
with $\beta = 1$, random intercepts $\gamma_i$, $\tau = 0.5$ and only one cluster-constant covariate $x_1,\dots,x_{50}$. We intend to fit a mixed model of setup (\ref{eq_example}) without early stopping, i.e.\ we choose a very high number $m_\text{stop} = 5000$ of total iterations and let the algorithm converge to its limit with the original boosting package \texttt{mboost}. Averaged over 100 simulation runs, we obtain the coefficient estimates $\hat{\beta} = 1.00$ for \texttt{lme4} \citep{Bates.lme4} but only $\hat{\beta} = 0.85$ for \texttt{mboost}. This significant bias does evidently not arise from early stopping with respect to prediction but from the selection process itself. The effect $\beta$ is fitted to some extend until the algorithm eventually reaches some threshold and starts selecting the random intercepts. This shows two major disadvantages. The first is, \texttt{mboost} initially picks fixed effects as in a regular linear model and later adds the random structure, which prevents an organic updating process where fixed and random effects are fitted simultaneously. The second is, because of the simple structure of the random effects baselearner in \texttt{mboost}, the random effects estimates tend to be correlated with observed covariates once they are updated and thus the model produces biased estimates for fixed and random effects as well.




This constitutes a crucial malfunction whose removal is, amongst others, relevant for an adequate development of boosting methods for joint models \citep{Waldmann.2017}, where the random structure plays an important role in connecting longitudinal and time-to-event models.

To overcome these issues, we propose a new algorithm which combines successful concepts of both gradient and likelihood-based boosting \citep{Tutz.2006, Tutz.2007}. The algorithm uses the organic updating scheme of the likelihood-based boosting framework for generalized mixed models proposed in \citep{Groll.Diss, Tutz.2010}, where random effects are updated alongside fixed effects. In addition, it fits the random effects using an improved version of the fitting technique from \texttt{mboost} equipped with a specific correction matrix, which ensures uncorrelated random effect estimates and was originally proposed in \citep{Griesbach.2019}. We hence overcome the problems of gradient boosting in random effects estimation but keep all the advantages the algorithm has over other approaches.

The remainder of the paper is structured as follows: Section \ref{sec_methods} formulates the underlying model and the updated boos\-ting algorithm as well as a detailed discussion of the changes. The algorithm is then evaluated and compared using an extensive simulation study described in Section \ref{sec_simulation} and applied to predict new cases of SARS-CoV-2 infections in Section \ref{sec_data}. Finally, the results and possible extensions are discussed.

\section{Methods}\label{sec_methods}

\subsection{Model Specification}\label{ssec_model}
For clusters $i = 1,\dots,n$ with observations $j = 1,\dots,n_i$ we consider the linear mixed model
\begin{equation*}
	y_{ij} = \beta_0 + \bx_{ij}^T \bbet + \bz_{ij}^T \bgam_i + \varepsilon_{ij},
\end{equation*}
with covariate vectors $\bx_{ij}^T = (x_{ij1},\dots,x_{ijp})$ and $\bz_{ij}^T = \\(z_{ij1},\dots,z_{ijq})$ referring to the fixed and random effects $\bbet$ and $\bgam_i$, respectively. The random components are assumed to follow normal distributions, i.e. $\varepsilon_{ij} \sim \mathcal{N}(0, \sigma^2)$ for the model error and $\bgam_i \sim \mathcal{N}^{\otimes q}(\boldsymbol{0}, \bQ)$ for the random effects. This leads to a cluster-wise notation
\begin{equation*}
	\by_i = \beta_0 \bone + \bX_i \bbet + \bZ_i \bgam_i + \beps_i
\end{equation*}
with $\by_i = (y_{i1}, \dots, y_{in_i})^T$, $\bone = (1,\dots,1)$, $\bX_i = (\bx_{i1},\dots,\bx_{in_i})^T$, $\bZ_i = (\bz_{i1},\dots,\bz_{in_i})^T$ and $\beps_i = (\varepsilon_{i1},\dots,\varepsilon_{in_i})$. Finally, we get the common matrix notation
\begin{equation}\label{eq_full_model}
	\by = \beta_0 \bone + \bX \bbet + \bZ \bgam + \beps
\end{equation}
of the full model with observations $\by = (\by_1^T,\dots,\by_n^T)^T$, design matrices $\bX = [\bX_1^T,\dots,\bX_n^T]^T$ and the block-diagonal $\bZ = \dg(\bZ_1, \dots,\bZ_n)$. The random components $\beps = (\beps_1^T,\dots,\beps_n^T)^T$ and $\bgam = (\bgam_1^T,\dots,\bgam_n^T)^T$ have corresponding covariance matrices $\sigma^2 \boldsymbol{I}_N$ and $\dg(\bQ,\dots,\bQ)$ where $\boldsymbol{I}_N$ is the $N = \sum n_i$ dimensional unit matrix.

In order to perform inference, let $\btet = (\beta_0, \bbet^T, \bgam^T)$ denote the effects and $\bphi = (\sigma^2, \btau)$ information of the random structure, where $\btau$ contains the values of $\bQ$. The log-likelihood of the model is
\begin{equation*}
	\ell(\btet, \bphi) = \sum_{i = 1}^{n} \log \int f(\by_i|\btet, \bphi) p(\bgam_i|\bphi) d \bgam_i,
\end{equation*}
where $f(\cdot|\btet, \bphi)$ and $p(\cdot|\bphi)$ denote the normal densities of the model error and the random effects. Laplace approximation following  \citep{Breslow.1993} results in the penalized log-likelihood
\begin{equation}\label{eq_likelihood}
	\ell^\text{pen}(\btet, \bphi) = \sum_{i = 1}^{n} \log f(\by_i|\btet, \bphi) - \frac12 \sum_{i = 1}^{n} \bgam_i^T \bQ^{-1} \bgam_i,
\end{equation}
see e.g. \citep{Tutz.2010}.

For a gradient boosting framework we define the predictor functions $\eta_\beta(\bx) = \beta_0 + \bx^T\bbet$ for fixed effects and $\eta_{\pp\gamma i}(\bz) = \bz^T\bgam_i$ for the random structure yielding
\begin{equation*}
	y_{ij} = \eta_{\beta ij} + \eta_{\pp\gamma ij} + \varepsilon_{ij}
\end{equation*}
for a single observation with $\eta_{\beta ij} = \eta_\beta (x_{ij})$ and $\eta_{\pp\gamma ij} = \eta_{\pp\gamma i} (z_{ij})$. The full model (\ref{eq_full_model}) is
\begin{equation*}
	\by = \bet_\beta + \bet_{\pp\gamma} + \beps
\end{equation*}
with $\bet_\beta = (\eta_{\beta ij})_{ij}$ and $\bet_{\pp\gamma} = (\eta_{\pp\gamma ij})_{ij}$. The fixed effects predictor is separated into single baselearner functions
\begin{equation*}
	\eta_{\beta}(\bx) = h_1(x_1) + \dots + h_p(x_p),
\end{equation*}
where each baselearner $h_r(x_r) = \beta_{0r} + \beta_rx_r$ models the (linear) effect of one single covariate. The predictor for the random structure consists of one single baselearner
\begin{equation*}
	\eta_{\gamma}(\bz) = h_\gamma(\bz),
\end{equation*}
which is explained in detail further below. Finally, we define the loss function
\begin{equation}\label{eq_loss_function}
	\rho(y, \eta(x,z)) = \frac{1}{2}(y - \eta(x, z))^2
\end{equation}
as the regular quadratic loss for $\eta(x, z) = \eta_{\beta}(x) + \eta_{\pp\gamma}(z)$.

\subsection{Boosting Algorithm} The \grb algorithm minimizes the loss (\ref{eq_loss_function}) between predicted and observed values via componentwise gradient-boosting while simultaneously accounting for the random structure by maximizing w.r.t. the log-likelihood (\ref{eq_likelihood}). We first write the procedure in compact form and give a detailed description in the following subsection. Let $\mathcal{I} = \{ij\colon i=1,\dots,n, j=1,\dots,n_i\}$ denote the index set of all observations.

\noindent\rule[0.5ex]{\linewidth}{1pt}

\begin{titlemize}{\textbf{Algorithm} \grb}
	\item \textbf{Initialize} predictors $\hat{\bet}_{\pp}^{[0]} = \hat{\bet}_\beta^{[0]} + \hat{\bet}_{\gamma\pp}^{[0]}$ and specify proper baselearners $\hat{h}_{\beta 1}^{[0]},\dots,\hat{h}_{\beta p}^{[0]}$ and $\hat{h}_{\pp\gamma}^{[0]}$. Set $\hat{\sigma}^{2[0]}$ and $\hat{\btau}^{2[0]}$ and choose $m_\text{stop}$ and $\nu$.
	
	\item \textbf{for} $m=1$ to $m_\text{stop}$ \textbf{do}
	
	\item[] \textbf{step1: Update $\bet_\beta^{\phantom{Gg}}$}\\
	Compute the negative gradient vector $\bu^{[m]}$ as
	\begin{equation*}
		\bu^{[m]} = \left(u_{ij}^{[m]}\right)_{ij \in \mathcal{I}} = \left(y_{ij} - \hat{\eta}_{\beta ij}^{[m-1]} - \hat{\eta}_{\gamma ij}^{[m-1]}\right)_{ij \in \mathcal{I}}
	\end{equation*}
	and fit $\bu^{[m]}$ separately to every baselearner specified for the fixed effects components:
	\begin{equation*}
		\bu^{[m]} \xrightarrow{\text{baselearner}} \hat{h}_{\beta r}^{[m]}, \quad r \in \{1,\dots,p\}
	\end{equation*}
	Select the component
	\begin{equation*}
		r^* = \underset{r}\argmin \ \sum_{ij \in \mathcal{I}} \left(u_{ij}^{[m]} - \hat{h}_{\beta r}^{[m]}\right)^2
	\end{equation*}
	that best fits $\bu^{[m]}$ and update $\hat{\bet}_\beta^{[m]} = \hat{\bet}_\beta^{[m-1]} + \nu \hat{h}_{\beta r^*}^{[m]}$.
	
	\item[] \textbf{step2: Update $\bet_\gamma^{\phantom{Gg}}$}\\
	Compute the negative gradient vector $\bu^{[m]}$ as
	\begin{equation*}
		\bu^{[m]} = \left(u_{ij}^{[m]}\right)_{ij \in \mathcal{I}} = \left(y_{ij} - \hat{\eta}_{\beta ij}^{[m]} - \hat{\eta}_{\gamma ij}^{[m-1]}\right)_{ij \in \mathcal{I}}
	\end{equation*}
	and fit $\bu^{[m]}$ separately to the random effects baselearner adapted to the current variance structure:
	\begin{equation*}
		\bu^{[m]} \xrightarrow{\text{baselearner}} \hat{h}_{\pp\gamma}^{[m]}
	\end{equation*}
	Update $\hat{\bet}_{\gamma\pp}^{[m]} = \hat{\bet}_{\gamma\pp}^{[m-1]} + \nu \hat{h}_{\pp\gamma}^{[m]}$.
	
	\item[] \textbf{step3: Update random structure}\\
	Receive current estimates of the random structure
	\begin{equation*}
		\hat{\btau}^{[m-1]} \rightarrow \hat{\btau}^{[m]}
	\end{equation*}
	based on maximization of the underlying log-likelihood (\ref{eq_likelihood}).
	
	\item[] \textbf{end for}
	
	\item \textbf{Stop} the algorithm at the best performing $m_*$ with respect to quality of prediction. Return $\hat{\bet}_{\pp}^{[m_*]} = \hat{\bet}_\beta^{[m_*]} + \hat{\bet}_{\gamma\pp}^{[m_*]}$ as the final predictors with corresponding coefficient estimates $\hat{\beta}_0^{[m_*]}$, $\hat{\bbet}^{[m_*]}$, $\hat{\bgam}^{[m_*]}$ and random structure $\hat{\sigma}^{2[m_*]}$ and $\hat{\btau}^{[m_*]}$.
\end{titlemize}

\noindent\rule[0.5ex]{\linewidth}{1pt}

The implementation of the \grb algorithm as well as a cross validation routine is provided with the new \texttt{R} add-on package \grb whose source code is hosted openly on \texttt{http://www.github.com/cgriesbach/grbLMM}.

\subsection{Computational details of the \grb Algorithm}
We give a stepwise description of the computational details of the \grb algorithm. In general, the predictor function $\eta(x,z) = \sum_r h_{\beta r}(x) + h_{\pp\gamma}(z)$ is split into several baselearner functions modelling fixed and random effects of the single covariates. While in each iteration only the best performing fixed effects baselearner is updated in a component-wise procedure enabling variable selection, the random effects are modelled together by a single baselearner which is updated in a second step to ensure proper estimation of the random structure. Although $\hat{\eta}$, $\hat{h}_{\beta r}$ and $\hat{h}_{\pp\gamma}$ are technically functions, we use this notation to refer to the fitted values generated by those functions, i.e.
\begin{equation*}
	\hat{h}_{\beta r} = \left(\hat{h}_{\beta r}(x_{ijr})\right)_{ij\in\mathcal{I}} = \left(\hat{\beta}_{0r} + \hat{\beta}_rx_{ijr}\right)_{ij\in\mathcal{I}}
\end{equation*}
for the fixed effects baselearners. Fitting a baselearner $h$ is achieved by applying the current residuals $\bu$ to the corresponding hat matrix $S$ yielding
\begin{equation*}
	\hat{h} = S\bu.
\end{equation*}

\textbf{Specification of the fixed effects baselearners.} Setting up the fixed effects baselearners is straight forward. Since each baselearner takes the form $h_{\beta r}(x) = \beta_{0r} + \beta_r x$ it has the corresponding hat matrix
\begin{equation*}
	S_{\beta r} = \tilde{\bx}_r(\tilde{\bx}_r^T \tilde{\bx}_r)^{-1}\tilde{\bx}_r^T, \quad r \in \{1,\dots,p\},
\end{equation*}
with $\tilde{\bx}_r = (\bone, \bx_r)$ and $\bx_r$ denoting the $r$th column of $\bX$.

\textbf{Specification of the random effects baselearner.} The random effects predictor consists of one single baselearner $h_{\pp\gamma}(z)$ defined via its hat matrix
\begin{equation}\label{eq_S_gamma}
	S_\gamma^{[m]} = \bZ\bC(\bZ^T \bZ + \sigma^{2[m-1]}\bQ_b^{-1[m-1]})^{-1}\bZ^T
\end{equation}
with $\bQ_b^{-1[m]} = \dg(\bQ^{[m]},\dots,\bQ^{[m]})^{-1}$. In general, (\ref{eq_S_gamma}) is the hat matrix of the regular BLUP estimator for a mixed model containing only random effects using the current estimates of the random structure. In addition, the estimates are corrected using the correction matrix $\bC$ to sure that the random effects estimates $\hat{\bgam}$ are uncorrelated with any observed covariates and thus prevent biased coefficient estimates for $\bgam$ and $\bbet$. A derivation of the matrix $\bC$ can be found in the appendix.

\textbf{Starting values.} The predictors
\begin{equation*}
	\hat{\bet}_\beta^{[0]} = \hat{\beta}_0^{[0]} \bone + \bX \hat{\bbet}^{[0]}, \quad \hat{\bet}_{\pp\gamma}^{[0]} = \bZ \hat{\bgam}^{[0]}
\end{equation*}
are initialized by using proper starting values for the coefficient estimates $\hat{\beta}_0^{[0]}$, $\hat{\bbet}^{[0]}$ and $\hat{\bgam}^{[0]}$. Fixed effect estimates $\hat{\bbet}^{[0]} = \boldsymbol{0}$ are set to zero. The remaining parameters $\hat{\beta}_0^{[0]}$, $\hat{\bgam}^{[0]}$, $\hat{\sigma}^{2[0]}$ and $\hat{\btau}^{[0]}$ are extracted from an initial mixed model fit
\begin{equation}\label{eq_start_int_model}
	\by = \beta_0 \bone + \bZ \bgam + \beps
\end{equation}
containing only intercept and random effects.

\textbf{Updating variance-covariance-components.} The current longitudinal model error is obtained by
\begin{equation*}
	\hat{\sigma}^{2[m]} = \Var(\by - \hat{\bet}^{[m]}).
\end{equation*}
Estimation of the random effects covariance matrix is achieved using an EM-type algorithm based on posterior modes and curvatures \citep[p.~238]{Fahrmeir.2001} by updating
\begin{equation*}
	\hat{\bQ}^{[m]} = \frac1n \sum_{i=1}^n \left(\boldsymbol{F}_{i}^{-1[m]} + \hat{\bgam_i}^{[m]}\hat{\bgam}_i^{[m]T}\right)
\end{equation*}
with
\begin{equation*}
	\bF_i^{[m]} = - \frac{\partial^2 \ell^\text{pen}}{\partial\bgam_i\partial\bgam_i^T} = \frac{1}{\hat{\sigma}^{2[m]}}\bZ_i^T\bZ_i + \left(\hat{\bQ}^{[m-1]}\right)^{-1}
\end{equation*}
denoting the current curvatures of the random effects model.

\textbf{Stopping iteration.}Gradient-boosting approaches rely on cross validating the loss function $\rho$ to achieve early stopping and thus shrinkage and variable selection. To account for the grouping structure, we partition the data cluster-wise into $k$ fairly equal subsets to compute
\begin{equation}\label{eq_pred_risk}
	CV_k^{[m]} = \frac1k\sum_{l = 1}^k \frac1{N_l}\left\|\by_l - \bX_l \hat{\bbet}_{-l}^{[m]}\right\|_2^2
\end{equation}
for every iteration $m = 1,\dots, m_\text{stop}$, where $N_l$ observations $\by_l$, $\bX_l$ of one subset $l = 1,\dots,k$ are used to evaluate the estimates $\hat{\bbet}_{-l}^{[m]}$ after $m$ iterations based on the remaining data. Averaged over all $k$ folds we then obtain $m_*$ by
\begin{equation*}
	m_* = \underset{m = 1,\dots,m_\text{stop}} \argmin CV_k^{[m]}.
\end{equation*}

In addition we also include early stopping based on the AIC, which approximates model complexity using the trace of each iterations projection matrix $\bH^{[m]}$,
\begin{equation*}
	\hat{\bet}^{[m]} = \bH^{[m]}\by,
\end{equation*}
mapping observations $\by$ to the corresponding fitted values after $m$ iterations. $\bH^{[m]}$ can be derived iteratively from the projection matrices
\begin{equation*}
	\bS^{[m]} = \bI_N - (\bI_N - \nu S_{\pp\gamma}^{[m]})(\bI_N - \nu S_{\beta}^{[m]})
\end{equation*} 
of each single step, where $S_{\beta}^{[m]}$ is the hat matrix corresponding to the in iteration $m$ selected fixed effect. Following \citep[p.~494]{Buehlmann.2007} we compute
\begin{equation*}
	\bH^{[m]} = \bI_N - (\bI_N - \bS^{[m]})\dots(\bI_N - \bS^{[1]})(\bI_N - \bS^{[0]}),
\end{equation*}
where $\bS^{[0]}$ is the projection matrix of the initial model fit (\ref{eq_start_int_model}). The current degrees of freedom are then obtained as
\begin{equation*}
	\text{df}^{[m]} = \text{tr}(\bH^{[m]})
\end{equation*}
and we get a corrected version of the AIC by
\begin{equation*}
	\text{AIC}^{[m]} = \log\hat{\sigma}^{2[m]} + \frac{1 + \text{df}^{[m]}/N}{1 - (\text{df}^{[m]} + 2)/N}
\end{equation*}
as described in \citep{Hurvich.2002}. Again, $m_*$ is chosen to minimize the AIC, i.e.
\begin{equation}\label{eq_caicc}
	m_* = \underset{m = 1,\dots,m_\text{stop}} \argmin \text{AIC}^{[m]}.
\end{equation}

\begin{table*}
	\centering
	\makebox[\textwidth][c]{
		\begin{tabular}{clcccccccccccccc}
			\hline
			&&&\multicolumn{2}{c}{\texttt{lme4}}&&\multicolumn{2}{c}{\texttt{mboost}}&&&\grba&&&&\grbb&\\
			$\tau$&p&&$\text{mse}_{\bbet}$&$\text{mse}_{\tau}$&&$\text{mse}_{\bbet}$&f.p.&&$\text{mse}_{\bbet}$&$\text{mse}_{\tau}$&f.p.&&$\text{mse}_{\bbet}$&$\text{mse}_{\tau}$&f.p.\\
			\hline
			0.4 & 10 && 0.014 & 0.001 && 0.044 & 0.53 && 0.013 & 0.001 & 0.48 && 0.013 & 0.001 & 0.50 \\ 
			0.4 & 25 && 0.019 & 0.001 && 0.043 & 0.42 && 0.014 & 0.001 & 0.31 && 0.014 & 0.001 & 0.44 \\ 
			0.4 & 50 && 0.029 & 0.001 && 0.046 & 0.33 && 0.015 & 0.001 & 0.20 && 0.016 & 0.001 & 0.40 \\ 
			0.4 & 100 && 0.055 & 0.001 && 0.050 & 0.23 && 0.019 & 0.001 & 0.14 && 0.022 & 0.001 & 0.37 \\ 
			0.4 & 500 && \bmin & \bmin && 0.051 & 0.09 && 0.021 & 0.001 & 0.04 && 0.043 & 0.001 & 0.29 \\ 
			\hline\hline
			0.8 & 10 && 0.043 & 0.016 && 0.155 & 0.72 && 0.043 & 0.014 & 0.49 && 0.041 & 0.014 & 0.51 \\ 
			0.8 & 25 && 0.047 & 0.016 && 0.160 & 0.56 && 0.043 & 0.014 & 0.35 && 0.042 & 0.014 & 0.43 \\ 
			0.8 & 50 && 0.062 & 0.014 && 0.168 & 0.40 && 0.050 & 0.012 & 0.24 && 0.050 & 0.012 & 0.39 \\ 
			0.8 & 100 && 0.087 & 0.017 && 0.163 & 0.30 && 0.050 & 0.015 & 0.16 && 0.053 & 0.015 & 0.38 \\ 
			0.8 & 500 && \bmin & \bmin && 0.185 & 0.12 && 0.057 & 0.015 & 0.05 && 0.078 & 0.015 & 0.29 \\ 
			\hline\hline
			1.6 & 10 && 0.153 & 0.243 && 0.615 & 0.84 && 0.155 & 0.230 & 0.47 && 0.152 & 0.230 & 0.50 \\ 
			1.6 & 25 && 0.180 & 0.213 && 0.643 & 0.67 && 0.178 & 0.195 & 0.34 && 0.175 & 0.194 & 0.42 \\ 
			1.6 & 50 && 0.187 & 0.261 && 0.613 & 0.50 && 0.176 & 0.259 & 0.29 && 0.174 & 0.258 & 0.40 \\ 
			1.6 & 100 && 0.206 & 0.256 && 0.666 & 0.36 && 0.174 & 0.238 & 0.14 && 0.173 & 0.239 & 0.39 \\ 
			1.6 & 500 && \bmin & \bmin && 0.693 & 0.14 && 0.166 & 0.255 & 0.05 && 0.184 & 0.251 & 0.29 \\ 
		\end{tabular}
	}
	\caption{Results for $\text{mse}_{\bbet}$, $\text{mse}_{\tau}$ and false positives in the random intercepts setup.}
	\label{tab_sim_rint}
\end{table*}

\begin{table*}
	\centering
	\makebox[\textwidth][c]{
		\begin{tabular}{clcccccccccccc}
			\hline
			&&&\multicolumn{2}{c}{\texttt{lme4}}&&\multicolumn{2}{c}{\texttt{mboost}}&&\multicolumn{2}{c}{\grba}&&\multicolumn{2}{c}{\grbb}\\
			$\tau$&p&&$\text{mse}_{\sigma}$&$\text{mse}_{\bgam}$&&$\text{mse}_{\sigma}$&$\text{mse}_{\bgam}$&&$\text{mse}_{\sigma}$&$\text{mse}_{\bgam}$&&$\text{mse}_{\sigma}$&$\text{mse}_{\bgam}$\\
			\hline
			0.4 & 10 && 0.000 & 1.132 && 0.000 & 2.339 && 0.000 & 1.192 && 0.000 & 1.193 \\ 
			0.4 & 25 && 0.000 & 1.156 && 0.000 & 2.350 && 0.000 & 1.201 && 0.001 & 1.204 \\ 
			0.4 & 50 && 0.000 & 1.183 && 0.000 & 2.365 && 0.001 & 1.194 && 0.001 & 1.202 \\ 
			0.4 & 100 && 0.000 & 1.352 && 0.001 & 2.252 && 0.001 & 1.278 && 0.001 & 1.298 \\ 
			0.4 & 500 && \bmin & \bmin && 0.001 & 2.224 && 0.001 & 1.241 && 0.007 & 1.351 \\ 
			\hline\hline
			0.8 & 10 && 0.000 & 2.555 && 0.000 & 7.833 && 0.000 & 2.570 && 0.000 & 2.569 \\ 
			0.8 & 25 && 0.000 & 2.533 && 0.000 & 7.987 && 0.000 & 2.528 && 0.001 & 2.530 \\ 
			0.8 & 50 && 0.000 & 2.807 && 0.001 & 7.402 && 0.001 & 2.759 && 0.001 & 2.766 \\ 
			0.8 & 100 && 0.000 & 2.863 && 0.001 & 7.827 && 0.001 & 2.701 && 0.001 & 2.725 \\ 
			0.8 & 500 && \bmin & \bmin && 0.002 & 7.511 && 0.001 & 2.837 && 0.007 & 2.943 \\ 
			\hline\hline
			1.6 & 10 && 0.000 & 7.840 && 0.000 & 30.465 && 0.000 & 7.850 && 0.000 & 7.841 \\ 
			1.6 & 25 && 0.000 & 8.723 && 0.001 & 28.510 && 0.000 & 8.710 && 0.001 & 8.703 \\ 
			1.6 & 50 && 0.000 & 8.469 && 0.001 & 30.294 && 0.001 & 8.413 && 0.001 & 8.408 \\ 
			1.6 & 100 && 0.000 & 8.593 && 0.001 & 30.930 && 0.000 & 8.416 && 0.001 & 8.425 \\ 
			1.6 & 500 && \bmin & \bmin && 0.002 & 29.345 && 0.001 & 7.823 && 0.007 & 7.909 \\ 
		\end{tabular}
	}
	\caption{Results for $\text{mse}_{\sigma}$ and $\text{mse}_{\bgam}$ in the random intercepts setup.}
	\label{tab_sim_rint_gasi}
\end{table*}

\section{Simulations}\label{sec_simulation}
Primary goal of the simulation study is to check, whether the \grb algorithm provides an organic selection process with adequate stopping contrary to classical approaches based on gradient boosting and thus we focus on accuracy of estimates with additional evaluation of the variable selection properties. Therefore, we compare the algorithm to \texttt{mboost} as the current gradient boosting method for random effects. In addition, we compare the algorithm to the classical approach implemented in the \texttt{lme4} function of the \texttt{lme4} package \citep{Bates.lme4}. The \grb Algorithm is included in two versions. The first, \grba, achieves its optimal stopping iteration based on cross-validation (\ref{eq_pred_risk}). The second, \grbb, uses the corrected AIC as formulated in (\ref{eq_caicc}). The maximum amount of stopping iterations was set to $m_\text{stop} = 5000$ for \texttt{mboost} and $m_\text{stop} = 1000$ for the \grb versions.

\subsection{Random Intercepts}
For $i = 1,\dots,50$ and $j = 1,\dots,10$ we consider the setup
\begin{equation*}
	y_{ij} = \beta_0 + \beta_1 x_{i1} + \beta_2 x_{i2} + \beta_3 x_{ij3} + \beta_4 x_{ij4} + \sum_{r=5}^{p} \beta_rx_{ijr} + \gamma_{0i} + \varepsilon_{ij}
\end{equation*}
with values $\beta_0 = 1$, $\beta_1 = 2$, $\beta_2 = 4$, $\beta_3 = 3$, $\beta_4 = 5$ and $\beta_6 = \dots = \beta_p = 0$ for the fixed effects, $x_{ir}, x_{ijr} \sim \mathcal{N}(0,1)$ for the cluster-constant and cluster-varying covariates and $\gamma_{0i} ~ \sim \mathcal{N}(0, \tau^2)$, $\varepsilon_{ij} \sim \mathcal{N}(0, \sigma^2)$ for the random structure with $\sigma = 0.4$ and $\tau \in \{0.4, 0.8, 1.6\}$. The total amount of covariates is evaluated for the five different cases $p \in \{10, 25, 50,\\ 100, 500\}$ ranging from low to high dimensional setups.

For $\bbet = (\beta_0,\dots,\beta_p)^T$ we consider mean squared errors
\begin{alignat*}{2}
	&\text{mse}_{\bbet} := \|\bbet - \hat{\bbet}\|^2, \quad &&\text{mse}_{\bgam} := \|\bgam - \hat{\bgam}\|^2,\\
	&\text{mse}_\sigma := (\sigma^2-\hat{\sigma}^2)^2, \quad &&\text{mse}_\tau := (\tau^2-\hat{\tau}^2)^2
\end{alignat*}
as an indicator for estimation accuracy. Variable selection properties are evaluated by calculating the false positives rates, i.e.\ the rate of non-infor\-mative covariates being selected. False negatives did not occur and hence are omitted.

Table \ref{tab_sim_rint} depicts the results for $\text{mse}_{\bbet}$, $\text{mse}_{\bgam}$ and false positives. The fixed effects estimation accuracy is slightly higher for \texttt{lme4} in low dimensional setups but rapidly decreases with growing amount of covariates while the gradient boosting based approaches tend to be fairly stable in higher dimensional setups. However, \texttt{mboost} is clearly behind its competitors. Estimation of the random intercepts variance performs equally well for \texttt{lme4} and \grb, even though the \grb versions are slightly better in most cases. False positives are higher for \texttt{mboost} in every single case and with respect to variable selection properties the main difference between \grba and \grbb becomes visible: The AIC based stopping process in \grbb leads to later stopping and thus higher false positives rates in exchange for slightly lower errors of the fixed effects, which has been already discussed for regular linear models in \citep{Mayr.2012}.

The BLUP properties are evaluated using the random effects estimation accuracy which is depicted in Table \ref{tab_sim_rint_gasi}. Similarly to $\text{mse}_{\bbet}$, the results for $\text{mse}_{\bgam}$ are not too far off between \texttt{lme4} and the \grb approaches. Yet again, \texttt{lme4} is slightly worse in many cases while \texttt{mboost} is clearly outrun by its competitors.

\subsection{Random Slopes}
We now consider a slightly altered setup by adding random slopes for the two informative cluster-varying covariates, i.e.
\begin{align}\label{eq_rslp}
	\begin{split}
		y_{ij} &= \beta_0 + \beta_1 x_{i1} + \beta_2 x_{i2} + \beta_3 x_{ij3} + \beta_4 x_{ij4} + \sum_{r=5}^{p} \beta_rx_{ijr}\\
		&\quad + \gamma_{0i} + \gamma_{1i}x_{ij3} + \gamma_{2i}x_{ij4} + \varepsilon_{ij}
	\end{split}
\end{align}
with
\begin{equation*}
	(\gamma_{0i}, \gamma_{1i}, \gamma_{2i}) \sim \mathcal{N}^{\otimes 3}(\boldsymbol{0}, \bQ), \quad \bQ := \begin{pmatrix}
		\tau^2 & \tau^* & \tau^*\\
		\tau^* & \tau^2 & \tau^*\\
		\tau^* & \tau^* & \tau^2
	\end{pmatrix},
\end{equation*}
where $\tau \in \{0.4, 0.8, 1.6\}$ and $\tau^*$ is chosen so that $\text{cor}(\gamma_{ki}, \gamma_{li}) = 0.6$ for all $k,l = 1,2,3$ holds. We evaluate the mean squared errors
\begin{alignat*}{2}
	&\text{mse}_{\bbet} := \|\bbet - \hat{\bbet}\|^2, \quad &&\text{mse}_{\bgam} := \|\bgam - \hat{\bgam}\|^2,\\
	&\text{mse}_\sigma := (\sigma^2-\hat{\sigma}^2)^2, \quad &&\text{mse}_{\bQ} := \|\bQ-\hat{\bQ}\|_F^2
\end{alignat*}
with $\|\cdot\|_F$ denoting the Frobenius norm of a given matrix.

Results for the random slopes setup are depicted in Tables \ref{tab_sim_rslp} and \ref{tab_sim_rslp_gasi}. In general, the relations are quite similar to the random intercepts setup, i.e. \texttt{lme4} and the \grb versions are not too far off each other while still \grb produces slightly better results for $\text{mse}_{\bbet}$ and $\text{mse}_{\bQ}$ in most of the cases. In the case of $\text{mse}_{\bgam}$, however, this does not hold and here \texttt{lme4} mainly outperforms \grb, while the estimation accuracy of \grb lies still in the same range.

\begin{table*}
	\centering
	\makebox[\textwidth][c]{
		\begin{tabular}{clcccccccccccccc}
			\hline
			&&&\multicolumn{2}{c}{\texttt{lme4}}&&\multicolumn{2}{c}{\texttt{mboost}}&&&\grba&&&&\grbb&\\
			$\tau$&p&&$\text{mse}_{\bbet}$&$\text{mse}_{\tau}$&&$\text{mse}_{\bbet}$&f.p.&&$\text{mse}_{\bbet}$&$\text{mse}_{\tau}$&f.p.&&$\text{mse}_{\bbet}$&$\text{mse}_{\tau}$&f.p.\\
			\hline
			0.4 & 10 && 0.018 & 0.009 && 0.081 & 0.71 && 0.020 & 0.013 & 0.46 && 0.020 & 0.013 & 0.42 \\ 
			0.4 & 25 && 0.025 & 0.009 && 0.086 & 0.65 && 0.022 & 0.012 & 0.27 && 0.021 & 0.012 & 0.36 \\ 
			0.4 & 50 && 0.039 & 0.010 && 0.087 & 0.57 && 0.023 & 0.012 & 0.18 && 0.024 & 0.012 & 0.33 \\ 
			0.4 & 100 && 0.071 & 0.010 && 0.092 & 0.48 && 0.025 & 0.012 & 0.10 && 0.028 & 0.012 & 0.31 \\ 
			0.4 & 500 && \bmin & \bmin && 0.104 & 0.22 && 0.027 & 0.011 & 0.03 && 0.050 & 0.011 & 0.24 \\ 
			\hline\hline
			0.8 & 10 && 0.058 & 0.127 && 0.298 & 0.77 && 0.072 & 0.124 & 0.44 && 0.069 & 0.123 & 0.42 \\ 
			0.8 & 25 && 0.069 & 0.128 && 0.297 & 0.75 && 0.073 & 0.121 & 0.28 && 0.071 & 0.121 & 0.36 \\ 
			0.8 & 50 && 0.082 & 0.129 && 0.302 & 0.71 && 0.074 & 0.119 & 0.17 && 0.074 & 0.119 & 0.33 \\ 
			0.8 & 100 && 0.122 & 0.106 && 0.311 & 0.65 && 0.078 & 0.094 & 0.11 && 0.082 & 0.095 & 0.31 \\ 
			0.8 & 500 && \bmin & \bmin && 0.344 & 0.38 && 0.082 & 0.124 & 0.04 && 0.102 & 0.125 & 0.25 \\ 
			\hline\hline
			1.6 & 10 && 0.227 & 1.922 && 1.149 & 0.58 && 0.280 & 1.829 & 0.41 && 0.273 & 1.828 & 0.42 \\ 
			1.6 & 25 && 0.240 & 1.925 && 1.138 & 0.58 && 0.277 & 1.808 & 0.29 && 0.270 & 1.807 & 0.36 \\ 
			1.6 & 50 && 0.269 & 1.578 && 1.144 & 0.55 && 0.294 & 1.435 & 0.19 && 0.289 & 1.440 & 0.32 \\ 
			1.6 & 100 && 0.285 & 1.895 && 1.171 & 0.54 && 0.299 & 1.852 & 0.14 && 0.292 & 1.853 & 0.32 \\ 
			1.6 & 500 && \bmin & \bmin && 1.357 & 0.36 && 0.320 & 1.804 & 0.04 && 0.330 & 1.809 & 0.25 \\ 
		\end{tabular}
	}
	\caption{Results for $\text{mse}_{\bbet}$, $\text{mse}_{\bQ}$ and false positives in the random slopes setup.}
	\label{tab_sim_rslp}
\end{table*}

\begin{table*}
	\centering
	\makebox[\textwidth][c]{
		\begin{tabular}{clcccccccccccc}
			\hline
			&&&\multicolumn{2}{c}{\texttt{lme4}}&&\multicolumn{2}{c}{\texttt{mboost}}&&\multicolumn{2}{c}{\grba}&&\multicolumn{2}{c}{\grbb}\\
			$\tau$&p&&$\text{mse}_{\sigma}$&$\text{mse}_{\bgam}$&&$\text{mse}_{\sigma}$&$\text{mse}_{\bgam}$&&$\text{mse}_{\sigma}$&$\text{mse}_{\bgam}$&&$\text{mse}_{\sigma}$&$\text{mse}_{\bgam}$\\
			\hline
			0.4 & 10 && 0.000 & 3.349 && 0.002 & 6.385 && 0.002 & 4.482 && 0.003 & 4.480 \\ 
			0.4 & 25 && 0.000 & 3.469 && 0.002 & 6.364 && 0.003 & 4.530 && 0.003 & 4.537 \\ 
			0.4 & 50 && 0.000 & 3.640 && 0.002 & 6.288 && 0.003 & 4.551 && 0.003 & 4.589 \\ 
			0.4 & 100 && 0.000 & 4.021 && 0.003 & 6.229 && 0.003 & 4.536 && 0.004 & 4.681 \\ 
			0.4 & 500 && \bmin & \bmin && 0.005 & 6.757 && 0.003 & 4.453 && 0.011 & 5.035 \\ 
			\hline\hline
			0.8 & 10 && 0.000 & 5.911 && 0.002 & 17.191 && 0.002 & 6.923 && 0.003 & 6.924 \\ 
			0.8 & 25 && 0.000 & 6.265 && 0.002 & 17.015 && 0.003 & 7.015 && 0.003 & 7.018 \\ 
			0.8 & 50 && 0.000 & 6.454 && 0.003 & 16.684 && 0.003 & 6.956 && 0.003 & 7.000 \\ 
			0.8 & 100 && 0.000 & 7.268 && 0.004 & 16.233 && 0.003 & 7.060 && 0.004 & 7.180 \\ 
			0.8 & 500 && \bmin & \bmin && 0.011 & 17.578 && 0.003 & 6.953 && 0.011 & 7.555 \\ 
			\hline\hline
			1.6 & 10 && 0.000 & 14.514 && 0.001 & 58.931 && 0.002 & 16.970 && 0.003 & 16.970 \\ 
			1.6 & 25 && 0.000 & 14.831 && 0.002 & 56.374 && 0.002 & 16.605 && 0.003 & 16.616 \\ 
			1.6 & 50 && 0.000 & 15.797 && 0.002 & 54.377 && 0.002 & 17.124 && 0.003 & 17.147 \\ 
			1.6 & 100 && 0.000 & 15.384 && 0.003 & 58.784 && 0.003 & 16.682 && 0.004 & 16.807 \\ 
			1.6 & 500 && \bmin & \bmin && 0.010 & 62.002 && 0.003 & 17.658 && 0.011 & 18.255 \\ 
		\end{tabular}
	}
	\caption{Results for $\text{mse}_{\sigma}$ and $\text{mse}_{\bgam}$ in the random slopes setup.}
	\label{tab_sim_rslp_gasi}
\end{table*}

\section{SARS-CoV-2 Data}\label{sec_data}
We illustrate the algorithm based on new cases of severe acute respiratory syndrome coronavirus 2 (SARS-CoV-2) infections in European countries over time. Aim of this illustration is to showcase the improved predictive quality and variable selection properties of  \texttt{grbLMM} as well as the biased random effects estimates of \texttt{mboost} which weaken its performance.

The data collects new infections up to 21st of June (2020) with several additional covariates and is publicly available\footnote{\texttt{https://github.com/owid/covid-19-data/tree/\\master/public/data}, accessed on June 24th, 2020}. Due to completion of data, we restricted the analysis to Europe only leading to $n = 22$ countries with $N = 2203$ measurements overall. Covariates included in the analysis are listed below.

\begin{itemize}
	\item 1: \texttt{time} - time in days
	\item 2: \texttt{str} - stringency index
	\item 3: \texttt{popd} - population density
	\item 4: \texttt{pop} - population
	\item 5: \texttt{age} - median age
	\item 6: \texttt{gdp} - gross domestic product per capita
	\item 7: \texttt{dia} - diabetes prevalence
	\item 8: \texttt{fsm} - female smokers
	\item 9: \texttt{msm} - male smokers
	\item 10: \texttt{beds} - hospital beds per thousands
	\item 11: \texttt{life} - life expectancy
\end{itemize}

The data set was split with ratio 2:1 into a training sample, on which the random intercept models
\begin{align}\label{eq_actg_mod}
	\begin{split}
		y_{ij} &= \beta_0 + \sum_{r = 1}^{11} \beta_r x_{ijr} + \gamma_{0i} + \varepsilon_{ij}
	\end{split}
\end{align}
were fit, and a test sample used for evaluating the predictive performance by calculating the mean squared prediction error. 

\begin{figure*}
	\centering
	\makebox[\textwidth][c]{
		\includegraphics[width = 150mm, height = 80mm]{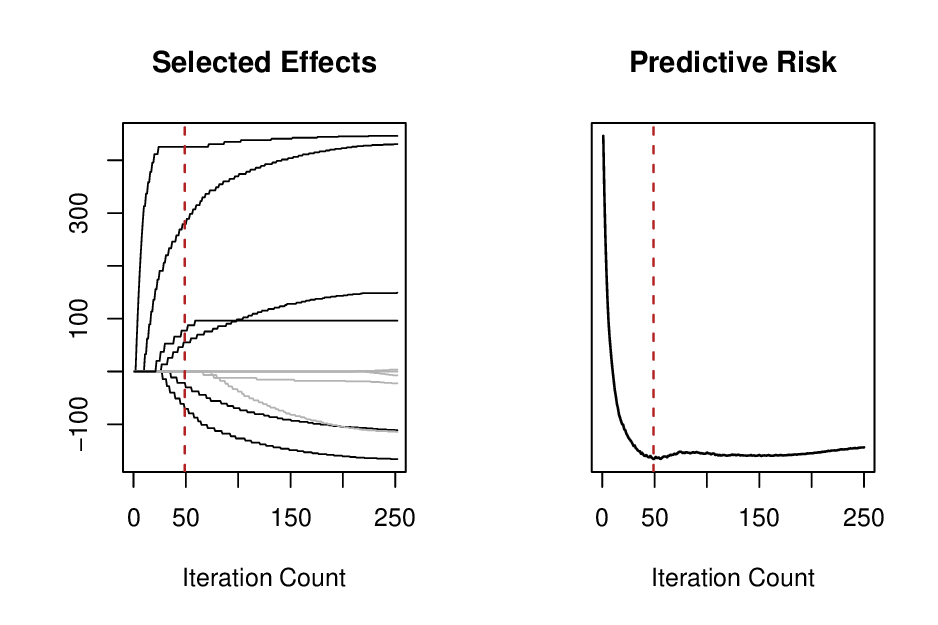}
	}
	\caption{Coefficient paths per iteration count for \texttt{grbLMM} with corresponding predictive risk. Optimal stopping iteration is at $m_* = 49$ and thus six covariates are excluded from the final model.}
	\label{fig_coef}
\end{figure*}

Based on 10-fold cross validation, \texttt{mboost} determined $m_* = 491$ and \texttt{grbLMM} $m_* = 49$ as their best performing stopping iteration. Results of the corresponding model fits are depicted in Table \ref{tab_data_results}. It is evident, that \texttt{grbLMM} has a better prediction than its two competitors. For \texttt{grbLMM}, the well known coefficient progression of boosting approaches alongside the change of predictive power per iteration count are shown in Figure \ref{fig_coef}.

The variables \texttt{popd}, \texttt{gdp}, \texttt{fsm}, \texttt{msm} and \texttt{beds} did not get selected by \grb and thus $\hat{\beta}_3 = \hat{\beta}_6 = \hat{\beta}_8 = \hat{\beta}_9 = \hat{\beta}_{10} = 0$ whereas the remaining coefficient estimates mostly received varying amounts of shrinkage in relation to the maximum likelihood estimates and hence offer increased quality of prediction.

\begin{table}[h]
	\centering
	\begin{tabular}{c|cccc}
		& $\hat{\beta}_0$ & $\hat{\beta}_1$(\texttt{time}) & $\hat{\beta}_2$ (\texttt{str})& $\hat{\hat{\beta}}_3$ (\texttt{popd}) \\
		\hline
		\texttt{lme4} & 329.78 & -175.60 & 443.08 & 21.33 \\ 
		\grb & 346.09 & -70.78 & 289.16 & 0.00 \\ 
		\texttt{mboost} & 313.47 & -172.58 & 440.18 & 0.00 \\ 
		\hline\hline
		& $\hat{\beta}_4$ (\texttt{pop})& $\hat{\beta}_5$ (\texttt{age})& $\hat{\beta}_6$ (\texttt{gdp})& $\hat{\beta}_7$ (\texttt{dia}) \\
		\hline
		\texttt{lme4} & 464.77 & 141.68 & -30.66 & -130.06 \\ 
		\grb & 425.34 & 55.06 & 0.00 & -30.05 \\ 
		\texttt{mboost} & 427.23 & 38.08 & 0.00 & -36.55 \\ 
		\hline\hline
		& $\hat{\beta}_{8}$ (\texttt{fsm})& $\hat{\beta}_{9}$ (\texttt{msm})& $\hat{\beta}_{10}$ (\texttt{beds})& $\hat{\beta}_{11}$ (\texttt{life}) \\
		\hline
		\texttt{lme4} & -133.61 & -49.05 & 2.12 & 57.31 \\ 
		\grb & 0.00 & 0.00 & 0.00 & 77.55 \\ 
		\texttt{mboost} & 0.00 & -19.99 & 0.00 & 51.80 \\
		\hline\hline
		& $\hat{\sigma}$ & $\hat{\tau}$ & \multicolumn{2}{c}{$\log\text{mspe} \ (\text{sd})$} \\
		\hline
		\texttt{lme4} & 765.00 & 144.75 & 11.35 & (2.16) \\ 
		\grb & 793.92 & 101.44 & 10.47 & (2.49) \\ 
		\texttt{mboost} & 760.93 & $210.39^*$ & 11.24 & (2.39) \\
	\end{tabular}
	\caption{Results for \texttt{lme4}, \texttt{mboost} and \texttt{grbLMM}. Please note, that \texttt{mboost} does not return any estimate for $\tau$ and we used sd($\hat{\bgam}_0)$ as a makeshift estimate for comparison.}
	\label{tab_data_results}
\end{table}

\begin{figure}[h]
	\centering
	\includegraphics[scale = .8]{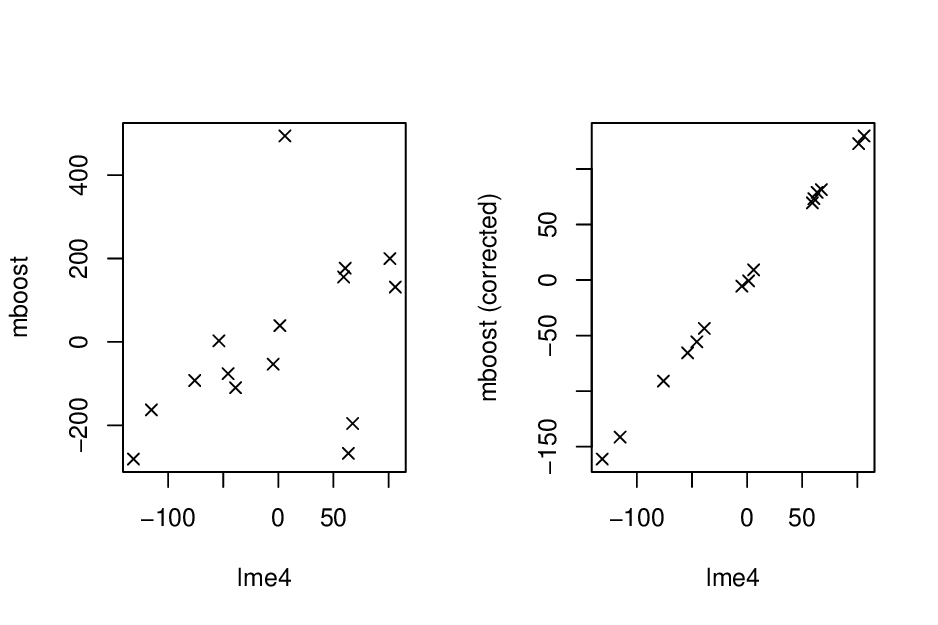}
	\caption{Comparison of random intercepts estimates for \texttt{lme4} and \texttt{mboost}. After counting out any correlations with observed covariates (right plot), the estimates are almost identical.}
	\label{fig_ranef}
\end{figure}

Although $\texttt{mboost}$ does not include 5 of the 11 candidate variables, the values fitted by \texttt{mboost} are almost identical to the ones returned by \texttt{lme4} (perfect Pearson correlation with a slope of 1.00). The reason for this is the higher spread of random effects, which is also reflected by the different values for $\hat{\tau}$. This leads to a biased selection process as effects of cluster-constant covariates are compensated by the random intercepts, while the actual fixed effects receive shrinkage. The random effects returned by \texttt{mboost} correlate with observed covariates and thus yield unjustified variable selection and potential overfitting, which is also indicated by the relatively small value for $\hat{\sigma}$. See Figure \ref{fig_ranef} for a visual representation.

\section*{Outlook and Discussion}
Due to its modified selection process and improved random effects baselearner, the proposed algorithm solves the issues evolving from the current approach implemented in the \texttt{mboost} package while simultaneously preserving the well known advantages of boosting approaches like implicit variable selection, improved quality of prediction and the possibility to deal with high dimensional data sets.

Still one major drawback in comparison to the concept implemented in \texttt{mboost} is that the proposed algorithm does not allow for model choice regarding the random structure, as the random effects have to be specified in advance and are not subject to any selection process. This, however, is only a minor issue, since the amount of different choices for the random structure is usually limited in real world applications and can also be evaluated afterwards using appropriate information criteria. Nevertheless it remains an interesting question, how we can construct a selection scheme for the random effects without losing the advantages gained by the \grb algorithm.

Due to the modular structure of the presented algorithm the authors are confident that extensions to more flexible models can be incorporated without many problems. This holds for both: changes in the predictor, such as non-linear effects based on P-splines \citep{Eiler.1996}, and changes in the outcome, such as a generalisation to non Gaussian outcomes in generalized additive regression models or even models for more than one parameter such as generalized additive models for location, shape and scale \citep{Rigby.2005}. Both are well established in current boosting frameworks \citep{Mayr.2012b} and it can be assumed, that those concepts can also be adapted to the \grb algorithm proposed in the present work.

\bibliographystyle{amsplain}
\bibliography{bibliothekus.maximus}

\clearpage
\appendix
\section*{Appendix}

\subsection*{Formulating the correction matrix $\bC$}
Due to the updating procedure, random effects estimates $\hat{\bgam}$ need to be corrected in order to ensure uncorrelated estimates with any other given covariates and thus also unbiased coefficient estimates $\hat{\bbet}$ for the fixed effects. At first, sets of covariates $\bX_{\text{c}s}$, $s=1,\dots,q$, have to be specified for each random effect which has to be corrected. For random intercepts $\bX_{\text{c}s}$ will include a column of ones as well as one representative of every cluster-constant covariate. For random slopes $\bX_{\text{c}s}$ will just contain a column of ones (which simplifies to centering the corresponding random effect) or additional cluster-constant covariates, if interaction effects are included for the covariate, the given random slope is specified for. The single correction matrices can then be computed by
\begin{equation*}
	\bC_s = \bX_{\text{c}s}(\bX_{\text{c}s}^T\bX_{\text{c}s})^{-1}\bX_{\text{c}s}, \quad s = 1,\dots,q
\end{equation*}
and one obtains the block diagonal $\tilde{\bC} = \dg(\bC_1,\dots,\bC_q)$. The final correction matrix $\bC$ is then obtained with
\begin{equation*}
	\bC = \bP^{-1}(\bI_{nq} - \tilde{\bC})\bP,
\end{equation*}
where $\bP$ is a permutation matrix mapping $\bgam$ to
\begin{equation*}
	\bP\bgam = \tilde{\bgam} = (\tilde{\bgam}_{1},\dots,\tilde{\bgam}_{q})
\end{equation*}
with $\tilde{\bgam}_s = (\gamma_{s1},\dots,\gamma_{sn})$. The product $\bC\bgam$ corrects each random effect $s$ for any covariates contained in the corresponding matrix $\bX_{\text{c}s}$ by counting out the orthogonal projections of the $s$th random effect estimates on the subspace generated by the covariates $\bX_{\text{c}s}$. This ensures the coefficient estimate for the random effects to be uncorrelated with any observed covariate.
\end{document}